\documentclass[usegraphicx,usedcolumn,useAMS,usenatbib]{mn2e}
\usepackage{graphics}
\usepackage{subfigure}
\usepackage{graphicx}
\usepackage{amssymb}
\usepackage{hyperref}

\title[Non-thermal processes in Abell 3376]
{Non-thermal processes in the cluster of galaxies Abell 3376}

\author[A. T. Araudo et al.]{Anabella T. 
Araudo$^{1,2,4}$\thanks{e-mail:aaraudo@fcaglp.unlp.edu.ar}, Sof{\'\i}a
A. Cora$^{2,3,4}$ and Gustavo E. Romero$^{1,2,4}$\\
$^1$Instituto Argentino de Radioastronom{\'\i}a (CCT La Plata, CONICET), 
C.C.5, 1894 Villa Elisa,  Buenos Aires, Argentina\\
$^2$Facultad de Ciencias Astron\'omicas y Geof{\'\i}sicas, Universidad Nacional
de La Plata, Paseo del Bosque, B1900FWA La Plata, Argentina\\
$^3$ Instituto de Astrof{\'\i}sica de La Plata (CCT La Plata, CONICET , UNLP), 
Observatorio Astron\'omico, Paseo del Bosque, B1900FWA \\La Plata, Argentina\\
$^4$ Consejo Nacional de Investigaciones Cient\'{\i}ficas y T\'ecnicas,
Rivadavia 1917, Buenos Aires, Argentina\\ 
}

\begin{document}

\date{}

\maketitle

\begin{abstract}

We model the high-energy emission that
results from  the interaction of relativistic particles with photons
and matter in the cluster of galaxies Abell 3376.
The presence of relativistic particles is inferred from
the recently found radio relics in this cluster, 
being one of the most prominent examples
of double opposite, giant ringlike radio structures.
Assuming that diffusive shock acceleration takes
place in the cluster regions where radio relics are observed, we calculate
the spectral energy distribution resulting from
the most relevant non-thermal processes, which are
synchrotron radiation, inverse Compton
scattering, relativistic Bremsstrahlung, and inelastic proton-proton
collisions. 
In the context of our model, the major
radiative component at high energies is inverse Compton scattering, 
which could reach
luminosities $L\sim 9\times 10^{41}$~erg~s$^{-1}$ in the energy range
between $\sim 1$~MeV and 10~TeV. Hadronic interactions would yield a minor
contribution to the overall non-thermal emission, but would dominate at
ultra-high energies. The cluster Abell~3376 might be
detectable at gamma-rays by HESS,
GLAST satellite and future planned Cherenkov arrays.
\end{abstract}

\begin{keywords}
cluster of galaxies: individual: Abell 3376 -- gamma-rays: theory --
radiation mechanisms: non-thermal
\end{keywords}

\section{Introduction}
\label{s_intro}

Clusters of galaxies are characterized 
by thermal emission in the $2$ to $10$~keV soft
X-ray region 
($L_{\rm X} \sim 10^{43} - 10^{45}\;\rm{erg\;s^{-1}}$)
as a result of Bremsstrahlung radiation
of the hot intracluster gas ($kT\sim 5-8$~keV).
However, the detection of radiation at radio wavelengths \citep{Feretti04},
as well as in the extreme ultraviolet ($0.07-0.4$~keV; EUV) \citep{Bowyer04}
and at hard X-rays ($20-80$~keV; HXR) \citep{fuscofemiano04}, 
indicates the presence of non-thermal activity in these systems.
There is general consensus that radio emission is produced by
synchrotron radiation originated from the interaction of
relativistic electrons with the cluster magnetic field
($B \sim \mu{\rm G}$) \citep{GovoniFeretti04}.
On the other hand, EUV radiation
might be produced either by a cooler thermal component ($kT\sim 2$~keV)
(\citealt{Lieu96}; \citealt{Mittaz98}) or by inverse Compton (IC) scattering of
cosmic microwave background (CMB) photons by
the same population of relativistic electrons responsible
for the radio emission (\citealt{EnsslinBiermann98}; \citealt{Lieu99}).
The latter process might also explain the HXR emission
\citep{FuscoFemiano99}, although this possibility is still matter of debate.

Regarding
the radio emission, whose non-thermal origin is firmly
established, 
clusters of galaxies present two different sources of diffuse large-scale
synchrotron emission,
known as `radio halos' and `radio relics'.
The former are located at the centre of clusters and are characterized 
by unpolarized radio
emission with structures that roughly resemble that shown by 
the X-ray emission,
whereas the latter are polarized radio sources
more irregularly shaped found at the
periphery of clusters (e.g. \citealt{FerettiGiovannini08}).
Radio relics have been observed in several clusters. The
most extended and powerful sources of this class have been detected 
in clusters with central radio halos, such as Coma 
\citep{Giovannini91} and Abell clusters A2163 \citep{Feretti01}, 
A2255 \citep{Feretti97}, 
A2256 \citep{Rottgering94} and A2744 \citep{Govoni01}. 
Only few clusters present double opposite relics, being the most
prominent examples those found in 
A3667 \citep{Rottgering97}, and A3376 \citet{Bagchi06}.
The large radio flux recently detected from the relics of A3376 
is an evidence of that relativistic leptons are present in those regions.
For this reason, in the present work we will be concerned with the 
non-thermal processes
taking place in the outskirts of the cluster Abell 3376 
where the two giant and almost symmetric radio relics are located.
Nevertheless, the model that we will present in this paper is also
valid for clusters which present only one relic. In the case of Abell 3376 
the radio flux is more than one order the magnitude larger than in other 
radio emitting clusters, like Abell 521 and Abell 2163.

There is increasing evidence that the radio emitting particles in relics are 
accelerated by accretion and merger shocks generated
in the intracluster medium 
(ICM) during cosmological large-scale structure formation.
Because of the electron short radiative lifetimes, radio
emission can be efficiently produced close to the location of the shock waves.
Thus, the origin of radio relics can be explained if they are
considered to trace the position of
these very large shocks (\citealt{Ensslin98}; \citealt{Hoeft04}).
According to the standard diffusive shock acceleration theory,
shock waves in the presence of even modest and 
turbulent magnetic fields
are sites of efficient acceleration of charged particles
\citep{Drury83}.
Another possibility may be
adiabatic compression in these environmental shock waves
\citep{EnsslinGopal01}.

Based on the inferred population of high energy particles involved in the
previously mentioned non-thermal processes,
it is natural to expect gamma-ray emission from galaxy clusters.
This emission could be generated from neutral
pion decay in cosmic ray collisions in the ICM \citep{VolkAha96},
or by IC scattering that involves CMB photons and relativistic electrons
\citep{AtoyanVolk00}.
Hydrodynamical cosmological simulations
are a useful tool to estimate the properties of the cosmic ray populations
in galaxy clusters and their effects on thermal cluster observables,
as well as to predict the
radiation emitted by these large virialized objects at $\gamma$-ray energies
(\citealt{Keshet03}, \citealt{Pfrommer07}, \citealt{Pfrommer08}).
A different numerical approach has been used by \citet{BerringtonDermer03}
to investigate the temporal evolution of particle and photon spectra
resulting from non-thermal processes at the shock fronts formed in merging
clusters of galaxies.
All these works point to clusters in general as sources
of $\gamma$-ray emission observable with GLAST satellite.

These theoretical results are consistent with the fact that no $\gamma$-ray 
emission has been detected so far with the current observational facilities,
as indicated by the lack of correlations between unresolved
EGRET $\gamma$-ray sources and nearby X-ray bright galaxy clusters
\citep{Reimer03}.
Only marginal evidence for
emission from A1758 within the location error contours
of the source 3 EG J1337+5029
was reported by \citet{Fegan05}.
Perseus and Abell 2029 galaxy clusters were observed by \citet{Perkins06}
with the Whipple 10 m Cherenkov telescope. They find no
evidence of point source or extended $\gamma$-ray emission in the TeV 
energy range. In addition, recent observations of the
clusters Coma and Abell 496
made by \citet{Domainko07} with the array of Chrerenkov telescopes HESS
have not detected significant signals of $\gamma$-ray emission
in exposure times of 
$\sim 10-20$ hours.

In the present work, we estimate the high energy emission 
of the recently detected radio relics in 
the nearby cluster Abell 3376. 
Their radio power at $1.4$~GHz, together with the assumption of 
an equipartition 
magnetic field, constrain effectively our model
which is specific of this particular cluster.
We assume a diffusive shock acceleration mechanism, 
and include different non-thermal
processes that may be acting in the radio structures.
We predict that the $\gamma$-ray photons created in this cluster 
should be detected in the near future by GLAST, as well as 
by the new generation of Cherenkov arrays in the southern hemisphere, 
such as HESS II.

The paper is organized as follows. Section \ref{s_cluster} describes
the main features of the cluster A3376.
Section \ref{s_accel} presents the acceleration and loss
mechanisms that affect the content of relativistic particles in the ICM.
The estimates of production of $\gamma$-rays and lower energy radiation
are given in Section \ref{s_emission}. Finally, 
we summarize our results and present our concluding remarks
in Section \ref{s_prospects}. 

\section{The cluster Abell 3376}
\label{s_cluster}

The rich cluster of galaxies Abell 3376 has been detected
by {\em ROSAT} and {\em XMM}-Newton through its X-ray emission 
revealing strong evidence for merger activity of subclusters.
It has a X-ray luminosity                                                
$L_{\rm X}(0.1-2.4\;\rm{keV}) \simeq 1.22\times 10^{44}\;h^{-2}\;\rm{erg\;s^{-1}}$.
This power is produced by thermal Bremsstrahlung radiation
from the hot ICM, which
has an overall temperature $T_{\rm X}\approx 5.8\times 10^7$~K
($\approx 5$~keV).
This cluster is located in the southern hemisphere 
($\alpha = 06^{\rm h} 0^{\rm m} 43^{\rm s}$, 
$\delta = -40^{\circ} 03^{\prime}$) at a 
redshift $z\approx 0.046$, which corresponds to a distance of
$d \sim 197$~Mpc for a standard $\Lambda$ cold dark matter ($\Lambda$CDM)
cosmology ($\Omega_{\rm m}$=0.3, $\Omega_{\Lambda}$=0.7, 
${\rm H}_{\rm o}= 100\, h \, {\rm km} \, {\rm s}^{-1} \, {\rm Mpc}^{-1}$,
with $h=0.7$).
The cluster mass has been estimated by applying
the virial theorem to the cluster member galaxies, assuming that mass
follows the galaxy distribution \citep{Girardi98}, giving a virial 
mass $M_{\rm vir} \sim 3.64\times 10^{14}\;h^{-1}\,M_{\odot}$.
The corresponding virial radius is $R_{\rm vir} \sim 0.98 \, h^{-1}$~Mpc.

Radio observations made by \citet{Bagchi06}
with the Very Large Array (VLA)  
instrument show the presence of two giant, ring-shaped 
structures in the periphery of the cluster at a distance 
of $\sim 0.7h^{-1}$~Mpc from its centre. 
The radio flux detected at $\nu = 1.4$~GHz is
$F_{\nu} = 302$~mJy, which corresponds to a radio luminosity of
$1.03\times 10^{40}\,h^{-2} \,{\rm erg}\,{\rm s}^{-1}$.
These structures have features typical of radio relics 
\citep{GiovanniniFerreti04}.
They fit quite well on a projected ellipse with
minor and major axis of $\sim 1.1h^{-1}$ and $\sim 1.4h^{-1}$~Mpc, 
respectively. 
Adopting a line-of-sight depth of $\sim 189~h^{-1}$~kpc, 
as \citet{Bagchi06}, 
the three-dimensional ellipsoid has a volume 
$V \sim 0.15~h^{-3}~{\rm Mpc}^3$.
From the composite map of radio and X-ray emissions 
shown by \citet{Bagchi06} (see their fig. 1a), we
infer that only $\sim 20\%$ of this volumen is filled by the radio relics
(i.e. $V_{\rm relic} \sim 0.003~h^{-3}~{\rm Mpc}^3$).

Some physical parameters of the cluster, relevant for the purpose of 
this work,
are not provided by the observations but can be obtained from simulations;
in concrete, the gas density in the relics, $n_{\rm H}$, and the shock 
velocity, $v_{\rm s}$.
We consider a cosmological non-radiative hydrodynamical {\em N}-body/SPH 
(Smoothed Particle Hydrodynamics) resimulation of a galaxy cluster 
that has been initially selected from a dark matter simulation for a
standard $\Lambda$CDM
cosmology with $\Omega_0 = 0.3$, $h = 0.7$, $\sigma_8 = 0.9$ and
$\Omega_b = 0.04$ \citep{Dolag05}.
The presence of radio relics in Abell 3376
might be interpreted as the result of an on-going merger of subclusters
\citep{Ensslin98}.
Thus, the chosen simulated cluster has similar 
dynamical state as the observed one;
it has a virial mass 
$M_{\rm vir} \sim 1.4 \times 10^{14}~h^{-1}~M_\odot$.
From the analysis of this simulation we obtain 
$n_{\rm H} \simeq  2\times 10^{-5}\,{\rm cm}^{-3}$ 
and $v_{\rm s} \simeq 1000\,{\rm km\;s^{-1}}$.
Considering that the temperature of the ICM, where the shock is 
propagating, is $T_{\rm ICM}\sim 10^{-1}T_{\rm relic}$ \citep{Hoeft04}
the Mach number results $M \sim 4.2$ (\citealt{Gabici}). For the rest 
of the paper we express all numerical values adopting $h=0.7$.

\section{Content of relativistic particles}
\label{s_accel}

Shock waves generated during the formation  and evolution of galaxy clusters
are the main source for thermalization of the intracluster gas
and the acceleration of particles \citep{Pfrommer06}.
The activity of radio galaxies embedded in the clusters also contribute
to the population of relativistic particles,
leaving fossil radio plasma that is detected as cavities
in X-ray surface brightness maps (e.g. \citealt{Churazov00}).
Two radio galaxies have been observed in the cluster Abell 3376.
The radio source MRC 0600-399 is associated with the 
second brightest cluster member galaxy, and the other radio source 
is possibly 
originated from an elliptical galaxy. These radio galaxies are
located within the central region of the cluster from where
thermal Bremsstrahlung emission is detected. Therefore, 
we can consider that the
ring-shape radio structures present in the periphery of the 
cluster are not connected with these point sources.

The morphology of the X-ray and radio emission observed in
the cluster Abell 3376
suggests that it is undergoing a merger.
As numerical simulations show (e.g. \citealt{Hoeft04}), 
shock waves propagate in both
directions along the line that connects the centres of the merging
clusters, with the radio relics observed almost exclusively
at the location of the shock fronts.
Thus, particle acceleration induced by
shock waves generated during this process is
a suitable scenario  for explaining the origin of the relics. 

Taking into account observational evidences and numerical
results, we assume that the content of relativistic particles
in the relics arises as a result of the acceleration by mergers shocks, 
neglecting the possible contribution of the radio galaxies.
In the following subsections, we describe the acceleration processes
and losses that affect both electrons and protons, which
determine the particle distributions and their subsequent evolution.

\subsection{Acceleration and losses}\label{acc_loss}

The diffuse radio emission produced at the location of relics
supports the presence of relativistic electrons and magnetic fields.
Although there is no observational evidence of the existence
of relativistic protons in the radio relics, these particles
could be accelerated in the same way as electrons are.

We focus here on the acceleration of particles at the (non-relativistic)
shock front via a diffusive process, such as first order Fermi mechanism. 
In this theory, the time-scale related to the rate of energy  
gain of particles is 
\begin{equation}
\tau_{\rm gain}= \frac{\gamma}{\dot\gamma_{\rm gain}} = 
\frac{\eta\; E}{e \, B\, c},
\label{eq1}
\end{equation}
where $\gamma$ is the Lorentz factor and 
$E$ is the energy  ($E=\gamma\,m\,c^2$)
up to which the particle, electron or proton,
is accelerated; $B$ is the magnetic field. The parameter
 $\eta = 2\pi f_{\rm{sc}} (c/v_{\rm s})^2$ involves the
velocity of the shock, $v_{\rm s}$, and the ratio $f_{\rm{sc}}$ between 
the mean free path of the particle and its gyroradius; the latter  
can be expressed as $r_{\rm g} =E[{\rm eV}]/(300\,Z\,B[\rm G])$,
being $Z$ the charge number of the particle.
For the shock velocity we assume a typical value  of 
$v_{\rm s} \sim 1000\, {\rm km}\,{\rm s}^{-1}$,
as it is inferred from the analysis of our cluster simulation,
being in agreement with
general simulations results (e.g. \citealt{Pfrommer06}).
Then, assuming Bohm diffusion (i.e. $f_{\rm{sc}} = 1$), 
we have $\eta = 5.7\times 10^{5}$. 

The rate of accelerated particles, $Q(\gamma)$,  
injected in the source follows a 
power-law energy distribution with a spectral index $\Gamma = 2.1$, 
typical of a diffusive acceleration mechanism 
and in accordance with the estimated Mach number $M$ (see \citealt{Blasi} 
for similar values of the spectral index in Coma cluster). 
Once the accelerated particles are injected,
their spectral energy distributions evolve 
as a consequence of radiative  
and advection losses. The latter refer to the escape 
of particles from the acceleration region.

In order to study the time-evolution of the energy distribution of particles,
we need to calculate the leptonic and hadronic radiative losses. 
In the following, we refer as primary particles to those relativistic
particles that have been accelerated at the shock front by the
Fermi-like mechanism. Secondary particles are the products originated 
from the interactions suffered by relativistic protons
with target thermal protons of the ICM. These inelastic proton-proton ($pp$)
collisions are the main channel of hadronic losses.  

The cooling rate of relativistic protons as a result
of $pp$ interactions can be estimated by
\begin{equation}
\dot{\gamma}_{pp} = -4.5\times10^{-16} n_{\rm H}\left[0.95 +
0.06\ln\left(\frac{\gamma}{1.1}\right)\right]\gamma\;\;\rm{s^{-1}}
\label{eqpp}
\end{equation}  
\citep{Mannheim94}, 
where $\gamma$ is the Lorentz factor for protons and $n_{\rm H}$ is the 
average hydrogen density of the relics obtained 
from the simulation. Part of the energy lost by
relativistic protons is used to create neutral pions which subsequently
decay into $\gamma$-rays. The rest is used in the creation of charged 
pions that will decay finally producing electron-positron 
pairs ($e^{\pm}$) and neutrinos.
The secondary pairs are cooled by the same radiative processes 
that affect primary electrons.
Lepton energy losses are estimated taking into account relativistic
Bremsstrahlung, IC interactions and synchrotron radiation 
(e.g. \citealt{GZ}). 

Since the ICM is fully ionized, we consider
Bremsstrahlung losses for bared nuclei, which are given by 
\begin{equation}
\dot{\gamma}_{\rm{Br}} = -6.9\times10^{-17} n_{\rm H} Z^2 
\left[\ln\gamma + 0.36\right]\gamma\;\;\rm{s^{-1}}.
\label{eqBremss}
\end{equation}      

Losses due to IC interactions depend 
on the energy density of photons, $U_{\rm ph}$. In the Thomson 
regime, they are given by
\begin{equation}
\dot{\gamma}_{\rm{IC}}= -3.2\times10^{-8} \,U_{\rm ph}\,\gamma^2
\;\;\rm{s^{-1}}.
\label{eqIC}
\end{equation}
The two photon fields relevant for the physical processes taking place
in clusters of galaxies are the CMB radiation and thermal X-ray photons
originated within the ICM itself through
thermal Bremsstrahlung.
The former is given by $U_{\rm CMB}=a\,T^4\,c^{-1}\,(1+z)^4$, where 
$a$ is the Stefan-Boltzmann
constant  and $T=2.725$~K is the CMB temperature; thus, 
$U_{\rm CMB} = 1.2\times 10^{-13}\,{\rm erg}\,{\rm cm}^{-3}$
at the cluster redshift.
The latter is expressed as 
$U_{\rm X}=L_{\rm X}\,(4\,\pi\,R^2\,c)^{-1}\,(1+z)^4$,
where $L_{\rm X}$ is the X-ray luminosity and $R$ is
the radius from which this luminosity is measured, which
is about $\sim 0.15-0.3$ per cent of the virial radius
\citep{balestra07}.
In the case of the cluster under study,  
$L_{\rm X} \simeq 2.44\times 10^{48} \,{\rm erg} \,{\rm s}^{-1}$ and 
$R \approx 0.3$ Mpc, giving
$U_{\rm X} \approx 1.3\times 10^{-15}\,{\rm erg}\,{\rm cm}^{-3}$. 
Since $U_{\rm X}$ is two orders of magnitude lower than $U_{\rm CMB}$, 
we can consider $U_{\rm X}$ negligible (in both Thomson and Klein-Nishina 
regimes of IC interactions) and
$U_{\rm ph} \sim U_{\rm CMB}$. 
Taking into account the energy for which
the emission of the CMB spectrum reaches its maximum 
($E_{\rm CMB} \sim 1.9\times10^{-4}$ eV),
leptons should reach energies greater than $1.6\times10^{15}$~eV
in order to satisfy the condition
$\xi = E_e E_{\rm CMB}/(m_ec^2)^2 \ge 1$,
that defines the Klein-Nishina regime of the IC interactions.
As we show in Section~\ref{MaxEner}, 
these high energies are never achieved by primary electrons; 
hence, IC scattering developes in Thomson regime.

Losses produced by synchrotron radiation are computed as 
\begin{equation}
\dot{\gamma}_{\rm{synch}}= -1.9\times10^{-9}B^2\,\gamma^2\;\;\rm{s^{-1}}.
\label{eqSynch}
\end{equation} 
The calculation of synchrotron losses requires the knowledge of  
magnetic field, $B$.
We estimate $B$ by assuming equipartition 
between the energy density of the 
magnetic field and the relativistic particles, that is,
\begin{equation}\label{equipartition}
\frac{B^2}{8\pi} = u_{e_1} + u_{p} + u_{e_2},  
\end{equation}
where $u_{e_1}$ and $u_{e_2}$ are the energy densities of primary 
electrons and secondary pairs, respectively,
and $u_p$ is the energy density of protons. 
The energy density of $i$-particles ($i= e_1, p, e_2$)  
is expressed as
\begin{equation}
\label{density}
u_{i}=\int_{E_i^{\rm min}}^{E_i^{\rm max}} E_i \; n(E_i)\; dE_i,
\end{equation}
where $n(E_i)$ is the particle energy distribution. 
For protons, $n(E_p)=K_p \,E_p^{-2.1}\,\exp(-E_p/E_p^{\rm{max}})$
but for leptons $n(E_{e_{1,2}})$ is a broken power-law:
\begin{equation}
n(E_{e})=\left\{ \begin{array}{ll}
K_e \,E_e^{-\Gamma_e} & E_e \leq E_{\rm b} \\
K_e^{\prime} \,E_e^{-\Gamma_e - 1}\,\exp(-E_e/E_e^{\rm{max}}) & 
E_e \geq E_{\rm b}.
\end{array}\right. 
\end{equation}
The spectral indices are $\Gamma_{e_1}=2.1$ for primary electrons and 
$\Gamma_{e_2}=2.08$ for secondary pairs; $E_{\rm b}$ is
the break energy (see Section \ref{particles} for the estimation 
of $E_{\rm b}$ and $K_e^{\prime}$). In all cases
$E_i^{\rm{max}}$ is the maximum energy that particles can reach, that
depends on the 
mechanism of energy gain and losses that affect the particles. 
The contribution of $E_i^{\rm{max}}$ to integral (\ref{density})
can be neglected because the slope of the particle distributions is 
negative: $n(E_{i}) \propto E_i^{-\Gamma_i}$.
On the other hand, we assume a minimum energy 
$E_{e_1}^{\rm min} = 1$~MeV for primary 
electrons and $E_p^{\rm min} = 1$~GeV for protons. Thus, for secondary
pairs we estimated $E_{e_2}^{\rm min} \sim E_p^{\rm min}/6\sim 0.17$~GeV.
Since the number of accelerated protons is unknown, we 
consider $u_p = a u_{e_1}$, where $a$ is 
a free parameter in our model. 
We adopt three different values: $a=0$ 
(no proton acceleration), $a=1$ (equal energy density in protons 
and in electrons), and  $a=100$  
(protons dominate the energy density budget). 

Efficient production of secondary electrons requires
a dense population of thermal protons as a target. However, this condition
is not found at the
peripheral cluster regions which are characterized by low density gas
($n_{\rm H} \sim  10^{-5}\,{\rm cm}^{-3}$). Therefore, 
the contribution of secondary pairs to
the total energy density (Eq.~\ref{equipartition}) is neglected.
Then, applying the constraint of the 
radio flux observed for Abell 3376 ($F_{\nu} = 302$~mJy), 
we estimate the magnetic field and the normalization constant
of the energy distribution of
primary accelerated electrons, $K_{e_1}$, and protons, $K_p$. 
Results are shown in Table \ref{t_1}.
Using the estimated spectral
energy distribution of protons, we obtain a ratio
between energy densities $u_{e_2}/u_{p}\sim 10^{-5}$,
which justifies our previous assumption about the secondaries.

\begin{table}
\caption{Magnetic field, $B$, and normalization 
constant of the energy distribution of primary accelerated 
electrons, $K_{e_1}$, and protons, $K_p$,
for the three different cases considered characterized 
by the parameter $a$.}
\label{t_1}
\begin{tabular}{lccc}
\hline
\hline
$a$ & $B$ & $K_{e_1}$ & $K_p$ \\
{}  & [G] & [erg$^{\Gamma - 1}$ cm$^{-3}$] & [erg$^{\Gamma - 1}$ cm$^{-3}$]\\
\hline
0    & $9\times 10^{-7}$  &  $1.4\times 10^{-15}$ & -\\
1    & $1.1\times 10^{-6}$  &  $1.1\times 10^{-15}$ & $8.7\times 10^{-16}$\\
100  & $3.4\times 10^{-6}$  &  $3\times 10^{-16}$ & $2.5\times 10^{-14}$\\
\hline
\end{tabular}
\medskip
\label{table1}
\end{table}

\begin{figure}
\includegraphics[width=0.9\hsize]{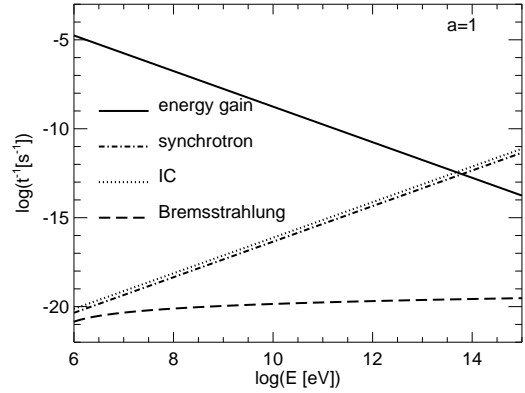} 
\caption{Rate of energy gain of primary electrons for 
$\eta \sim 5.7 \times 10^{5}$ compared with energy loss rates 
obtained for the case $a=1$.}
\label{losses}
\end{figure}

\subsection{Maximum energies}\label{MaxEner}

The theoretical maximum energy of primary particles is determined by the
competition between the rates of energy gain and losses,
as described in the previous section.
Fig.~\ref{losses} shows the time-scales of 
energy losses of primary electrons as a function of their energy
(Eqs.~\ref{eqBremss}, \ref{eqIC} and 
\ref{eqSynch}). These time-scales are compared with the
one corresponding to the acceleration rate, given by Eq.~\ref{eq1}.
From this comparison, we obtain the maximum energy allowed for electrons,
$E_{e_1}^{\rm{max}}$, 
which is $\sim 9\times 10^{13}$~eV.
For the case of protons, we take into account the energy loss rate
given by $pp$ interactions (Eq.~\ref{eqpp}), considering that
they have the same acceleration rate as electrons. 
We estimate proton maximum energies  
$E_p^{\rm{max}} \simeq 1.8\times 10^{21}$~eV and 
$4.6\times 10^{21}$~eV for cases with $a=1$ and $100$,
respectively.
At proton energies higher than $\sim 5\times10^{18}$~eV, photopair
production against CMB photons starts to be important, 
and beyond energies of
$\sim 5\times10^{19}$~eV photopion losses become dominant
(e.g. \citealt{Berezinsky88, kelner08}). However, in the current analysis
we do not need to take these losses into account since, as we will
see below, the actual maximum energy of protons in the relics
will be determined by additional constraints.

The values of maximum energy found for both electrons
and protons are valid as long as they 
allow the particles to remain within the acceleration region.
Thus, the particles have to satisfy the constraint 
$r_{\rm g} < l$, where 
$r_{\rm g}$ is the gyroradius of the particle
and $l$ is the size of the acceleration region.
By assuming that particles are accelerated by the shock waves
traced by the observed radio relics, we
adopt the approximation 
$l \simeq l_{\rm relic}\sim 0.3$~Mpc; this value is obtained from the
examination of the projected radio map of the cluster
Abell 3376. 
For the electrons with energies close to the maximum value  
$E_{e_1}^{\rm{max}}$, we find that $r_{\rm g}\sim 0.1$ pc.
Thus, the most energetic electrons are contained 
within the acceleration region.
The situation for protons is different
since the corresponding maximum energy, $E_p^{\rm{max}}$,
makes their gyroradius 
larger than $l_{\rm relic}$.
Hence, $E_p^{\rm{max}}$ is determined from the condition
$r_{\rm g}=l_{\rm relic}$.
This estimation gives $E_p^{\rm{max}} \simeq 9.7\times 10^{19}$ eV and
$2.6\times 10^{20}$ eV,
for $a=1$ and $100$, respectively. 
However, the time required  for protons to reach these energies
is higher than the lifetime of the relic 
($\tau_{\rm shock}^{\rm relic} \sim 1$ Gyr; see the 
next subsection). Thus,
the actual maximum energy to which protons can be accelerated is obtained
equating $\tau_{\rm acc}$ and $\tau_{\rm relic}$. 
The resulting values of $E_p^{\rm{max}}$ obtained from
this constraint are lower, being $\sim 5\times 10^{17}$ eV
and $\sim 1.3\times 10^{18}$ eV for $a=1$ and $100$, respectively.

The contribution of secondary pairs to the 
total spectral energy distribution (SED) will be taken into
account even though
the ambient conditions at the cluster outskirts
are not favorable for the generation of these particles 
through $pp$ interactions.
To this aim, we first calculate the injected spectrum of secondary pairs
taking into account the estimated
distribution of relativistic
protons, $n(E_p)$,
and the new parametrization of the inelastic cross-section of $pp$ 
interactions
\begin{equation}\label{cross-section}
\sigma_{\rm{inel}}(E_p) = 34.3 + 1.88L + 0.25L^2 \;\rm{mb},
\end{equation}
given by Kelner, Aharonian \& Vugayov (2006),
where $L = \ln(E_p/1\;\rm{TeV})$.
The maximum energies for secondary pairs 
are assumed to be equal to the maximum energy of the photons produced
by $\pi^0$-decay  (\citealt{kelner06}, Fig. 12).

\begin{table}[]
\begin{center}
\caption{Maximum energies obtained for primary electrons and protons
accelerated in the radio relics of Abell 3376 are given
in the first and second columns, respectively. The  maximum energy for 
secondary pairs $e^{\pm}$ is presented in the last column.}\label{Emax}
\begin{tabular}{cccc}
\hline
\hline
a  & $E_{e_1}^{\rm{max}}$ & $E_{p}^{\rm{max}}$ & $E_{e_2}^{\rm{max}}$ \\
{} & [eV]                 & [eV]             & [eV]                 \\
\hline
$0$   & $9\times10^{13}$ & -                  & -                   \\
$1$   & $9.3\times10^{13}$ & $5.0\times10^{17}$ & $4\times10^{16}$ \\ 
$100$ & $8.8\times10^{13}$ & $1.3\times10^{18}$ & $10^{17}$ \\
\hline
\end{tabular}
\label{table2}
\end{center}
\end{table}

We now analyse the time-evolution of the spectral energy distribution
of the three kind of relativistic particles present in the radio relics.

\subsection{Time-evolution of the energy distributions 
of relativistic particles}\label{particles}

Relativistic particles are injected in the radio
emitting region of the cluster with an energy distribution that follows
a power-law, which subsequently evolves as a consequence of
the radiative losses suffered by the particles.
Such evolution is estimated, as a first order approximation, 
by adopting a model in which the energy
distribution of the particles is determined by the following transport equation
\begin{equation}
\partial n\left(t,\gamma\right)/\partial t+\partial \dot{\gamma} 
n\left(t,\gamma\right)/\partial\gamma+n\left(t,\gamma\right)/
\tau_{\rm esc}=Q(t,\gamma)
\label{Ginz_equation}
\end{equation}
(e.g., \citealt{Khangulyan07}),
where $t$ is the time, 
$Q(t,\gamma)$ is the particle injection function, wich for 
present case will be time independent, and
$\tau_{\rm esc}$ is the advection escape
time defined as $\tau_{\rm esc} = l_{\rm{relic}}/v_{\rm{adv}}$.
In the shock rest-frame, the downstream (post-shock) plasma velocity
is given by $v_{\rm{adv}} \sim v_{\rm{shock}}/4$; thus, we
obtain $\tau_{\rm esc} \sim 1.1$~Gyr.

Particles are injected into the {\rm relics} with a rate $Q(\gamma)$
as the shock wave propagates
from the centre of the cluster up to the location of the relics.
For the purpose of the present work, the variable $t$
corresponds to different values of the
lifetime of the shock, $\tau_{\rm shock}$.
When the shock reaches the location of the observed radio relics,
this lifetime is referred to as $\tau_{\rm shock}^{\rm relic}$.
From the assumed shock velocity $v_{\rm s} \sim 1000\, {\rm km}\,{\rm s}^{-1}$
(Subsection~\ref{acc_loss}) and the position of the observed radio relics
in cluster Abell 3376, we have  
$\tau_{\rm shock}^{\rm relic}\sim 0.95$~Gyr. This value is consistent with the
one obtained from 
numerical simulations of cluster mergers  (e.g. \citealt{Hoeft04}).

The time-derivative $\dot{\gamma}$ is a function 
accounting for all the energy losses affecting leptons and protons, 
that is, mainly synchrotron and IC interactions in the former case and
$pp$ interactions in the latter one.
The solution of Eq.~(\ref{Ginz_equation}) for different shock lifetimes is
\begin{equation}
n(\gamma)=\frac{1}{|\dot{\gamma}|}\int\limits_{\gamma}^{\gamma_{\rm eff}} 
Q(\gamma'){\rm e}^{-\tau({\gamma,\gamma'})/\tau_{\rm esc}}
d\gamma',
\label{solution0}
\end{equation}
where $\gamma_{\rm eff}$ and  $\tau(\gamma,\gamma')$ are determined by 
\begin{equation}
\tau_{\rm shock}=\int\limits_{\gamma}^{\gamma_{\rm eff}}\frac{d\gamma'}{|\dot{\gamma'}|}
\;\;\;{\rm and}\;\;\;
\tau(\gamma,\gamma')=\int\limits_{\gamma}^{\gamma'}
\frac{d\gamma''}{|\dot{\gamma''}|}.
\label{ttt}
\end{equation}

\begin{figure*}
\centering
\includegraphics[width=0.4\textwidth]{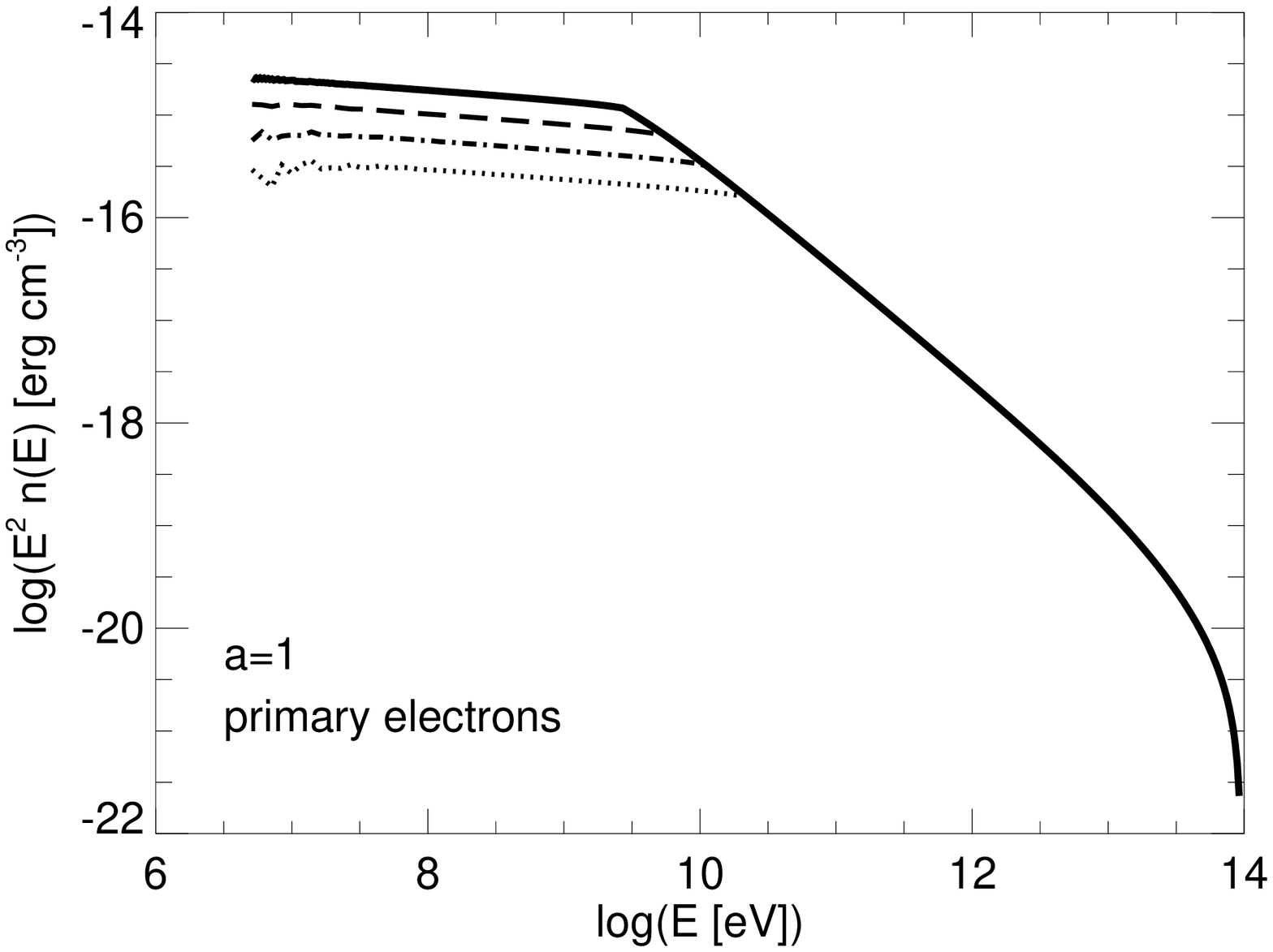}\qquad
\includegraphics[width=0.4\textwidth]{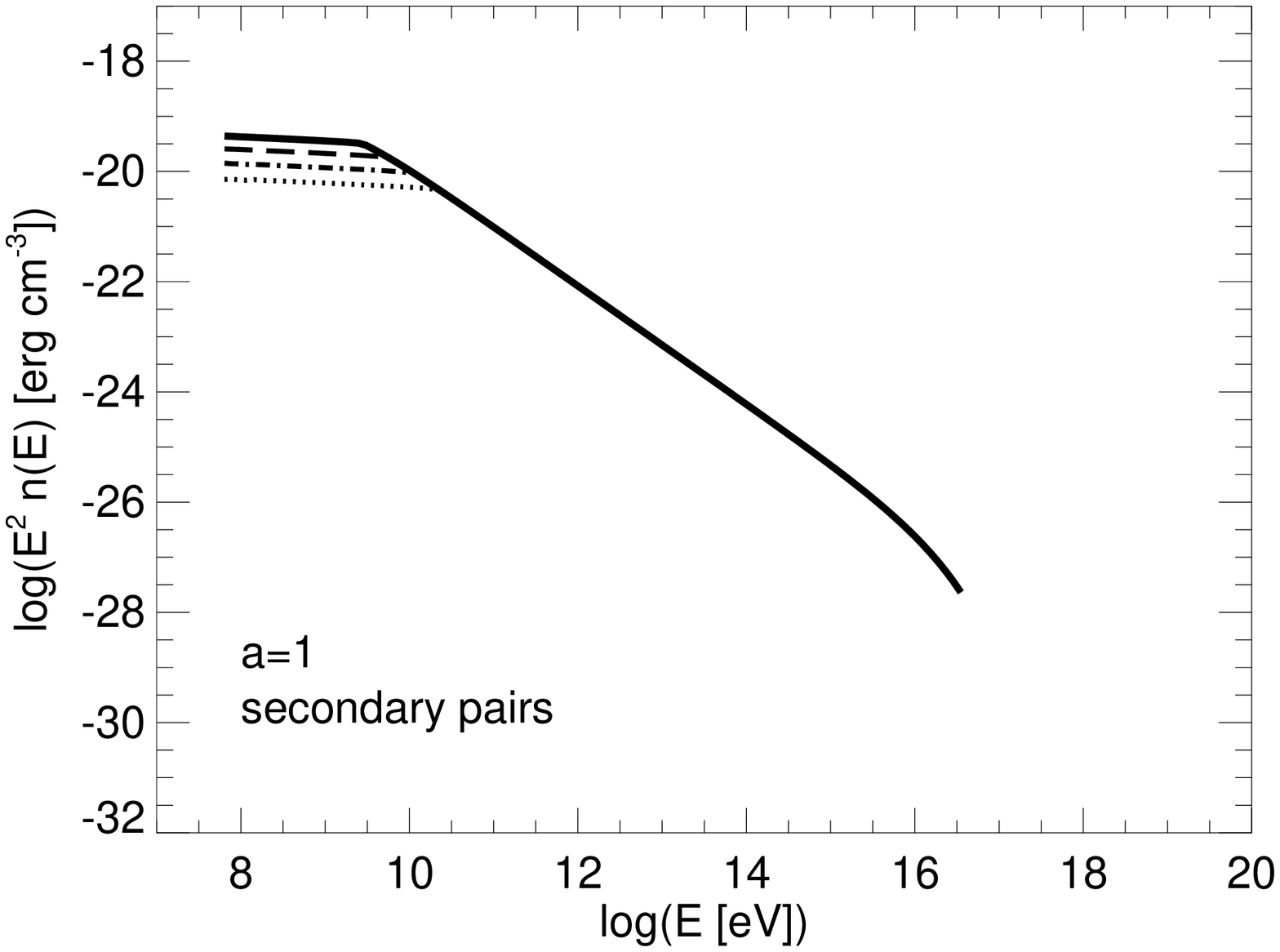}\\
\includegraphics[width=0.4\textwidth]{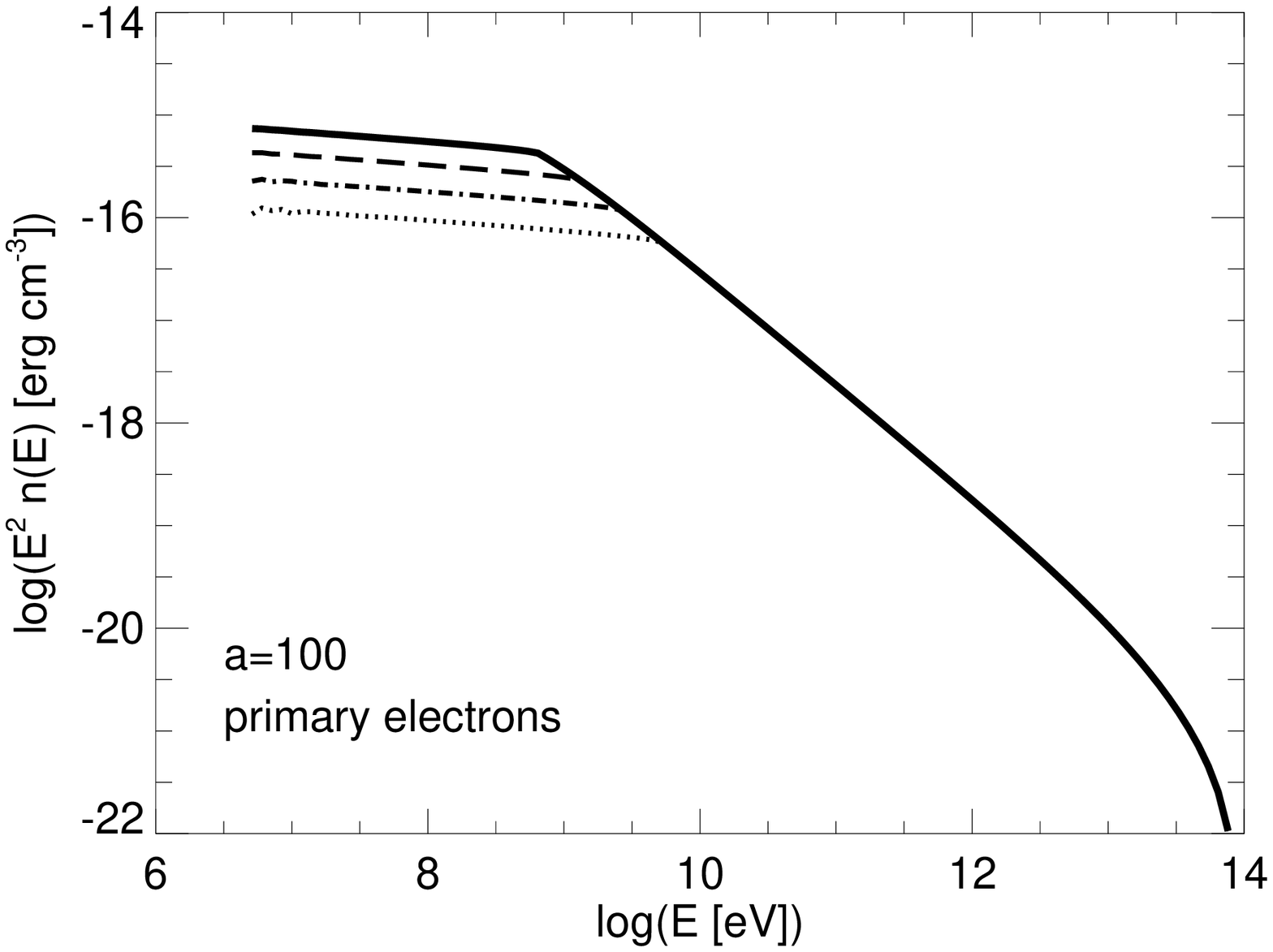}\qquad
\includegraphics[width=0.4\textwidth]{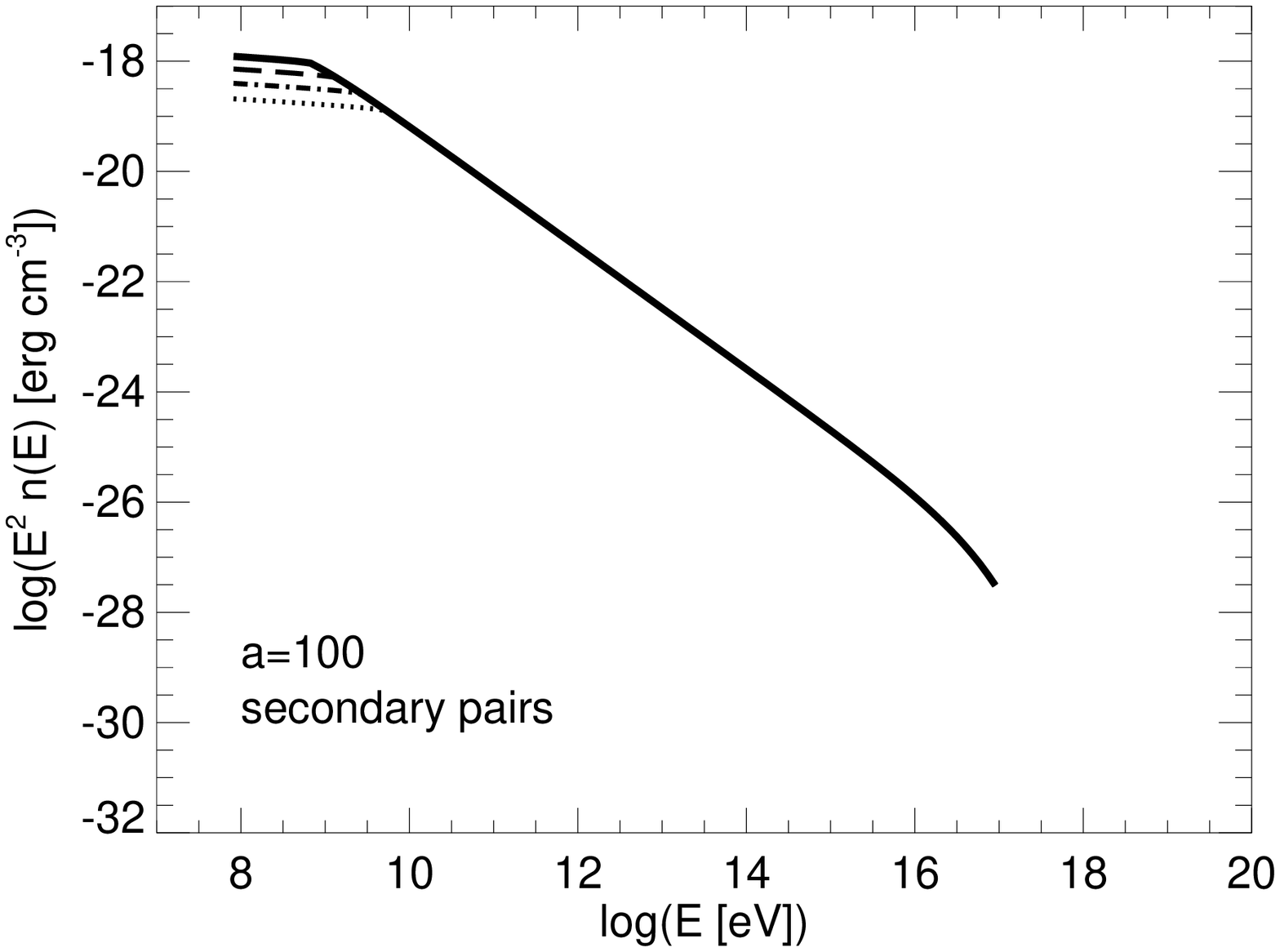}\\
\caption{Spectral energy distributions 
of primary electrons and secondary pairs for $a=1$ (upper panels) and
$a=100$ (lower panels).
Distributions for different particle injection time-scales 
($\tau_{\rm shock}$) are shown in each panel. Different curves correspond 
to the following values of
$\tau_{\rm shock}$: $0.125$~Gyr (dotted line), 
$0.25$~Gyr (dot-dashed line), $0.5$~Gyr
(dashed line) and $1$~Gyr (solid line). 
The steady regime is reached for ages
$\sim \tau_{\rm shock}^{\rm relic}$.
}
\label{evol}
\end{figure*}

The cooling time of leptons is 
$\tau_{\rm{cool}} = {\gamma}/{\dot\gamma_{\rm loss}} 
\sim 2\times 10^{20}\gamma^{-1}$~s.
If $\tau_{\rm{cool}} > \tau_{\rm{shock}}$ at a certain energy, the
particle energy distribution has the shape characterized by
$\Gamma_e=\Gamma$ and $K_e$ 
(see Table~\ref{table1}). There is a break
energy, $E_{\rm b}$, above of which $\tau_{\rm{cool}} < \tau_{\rm{shock}}$.
In this case, both primary electrons and secondary pairs
suffer  radiative losses that affect the injected particle spectrum.
As a consequence of this, the cooled spectrum has an index 
$\Gamma_{e_{1,2}}^{\prime} = \Gamma_{e_{1,2}} + 1$ with a normalization 
costant $K^{\prime}_{e_{1,2}} \sim E_{\rm b} K_{e_{1,2}}$  
for energies $E_{e_{1,2}} > E_{\rm b}$. 
The computed particle energy distributions of 
primary electrons and secondary pairs 
are shown in Fig.~\ref{evol}. 
As can be seen, the spectrum is broken at 
$E_{\rm b}\sim 5\times10^{9} - 5\times10^{10}$ eV
for values of $\tau_{\rm shock}$ comprised in the range $100$~Myr to $1$~Gyr,
being the break energies smaller for larger lifetimes of the shock.  

In this scenario, particles will reach the steady regime when
$\tau_{\rm{shock}} \geq \tau_{\rm{esc}}$. For this reason, as
$\tau_{\rm{shock}}^{\rm{relic}} \sim \tau_{\rm{esc}}$, we can consider
that by the time when the radio relic 
is observed, the particle energy distribution 
$n(\gamma)$ is already steady.

Regarding protons, the escape time-scale ($\tau_{\rm{esc}} \sim 1.1$~Gyr) 
and shock lifetime ($\tau_{\rm shock}^{\rm relic} \sim 1$ Gyr) are
shorter than the cooling time ($\tau_{pp} \sim 10^{3}$ Gyr). 
Consequently,
the energy distribution in the context of our scenario keeps the same 
spectral shape as that of the injected one.
Since $\tau_{\rm{esc}}$ is of the order of $\tau_{\rm shock}^{\rm relic}$, 
we can consider that the energy distribution of protons also reaches
the steady regime at $\tau_{\rm shock}^{\rm relic}$.

All these calculations give us the steady energy distribution of
the different population of relativistic particles, i.e.
primary electrons, secondary pairs,
and protons, that have become part of the relics.
Electrons are responsible for the radio emission of the cluster Abell 3376
observed in its giant ringlike radio structures. As we have
shown, this emission has been used to estimate the magnetic field
strength in the acceleration region, and the normalization and
maximum energies involved in the power-laws that represent the
particle energy distributions.
In the following section, we estimate the contribution of
both relativistic leptons and protons to the high energy
emission of the cluster, evaluating the detectability of
Abell 3376 in the MeV-TeV electromagnetic range.

\section{Production of gamma-rays and lower energy radiation}
\label{s_emission}

Using the steady distributions of particles estimated 
in the previous section, we calculate the SEDs
for the three different combinations of proton 
and lepton relativistic energy densities (i.e. cases $a = 0, 1, 100$).
We take into account the most relevant non-thermal radiative 
processes according 
to the conditions of the ambient medium in the cluster Abell 3376.

\subsection{Leptonic emission}
\label{s_leptonic}

The differential emissivities $q_{\gamma}(E_{\gamma})$ produced by 
leptons for synchrotron radiation, IC scattering and 
relativistic Bremsstrahlung
are calculated using the standard formulae
given by \citet{Blumenthal70} and \citet{Pacholczyk70}.  
The luminosity produced in the relics is given by the equation

\begin{equation}
E_{\gamma}L_{E_{\gamma}}=E_{\gamma}^{2}\; q_{\gamma}(E_{\gamma})\; 
V_{\rm{relic}},
\label{luminosity}
\end{equation}
where $E_{\gamma}$ is the photon energy and $V_{\rm{relic}}$ is
the volume of the emitting region. For the latter, we adopt a value of
$\sim 0.09\,\rm{Mpc^3}$,
as we have mentioned in Section~\ref{s_cluster}.
In the cases where a population of accelerated protons
is assumed to be present ($a=1, 100$), we estimate the contribution
of secondary pairs to the SEDs for completitude.

\begin{figure}
\includegraphics[width=0.9\hsize]{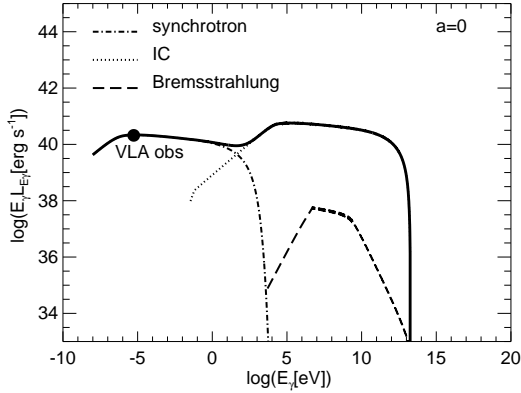} %
\caption{SED for a pure leptonic case ($a=0$), represented by a solid line.
The contribution of different radiative processes are shown:
synchrotron radiation (dot-dashed line), IC scattering (dotted line) and
relativistic Bremsstrahlung (dashed line). The VLA observation at $1.4$~GHz
is also included.}
\label{SED_0}
\end{figure}

Results of the luminosity produced in the case where only  primary
electrons are accelerated, characterized by $a = 0$, 
are shown in Fig. \ref{SED_0}. 
We can see that IC interactions are
the dominant process, with a luminosity 
$L_{\rm{IC}} \sim  9.1\times10^{41} \, {\rm erg} \, {\rm s}^{-1}$, 
at energies $E_{\gamma} \gtrsim 0.1$~MeV
and with a cut-off at $\sim 10$~TeV. The luminosity produced by 
relativistic Bremsstrahlung is negligible, in agreement with results
shown in Fig.~\ref{losses}. 

\subsection{Hadronic emission}
\label{s_hadronic}

Proton-proton interactions produce $\gamma$-ray emission through
$\pi^0$-decay.
The corresponding differential $\gamma$-ray emissivity 
is calculated as
\begin{equation}
q_{\gamma}(E_{\gamma})=2\int^{\infty}_{E_{\pi}^{\rm min}(E_\gamma)}
\frac{q_{\pi^0}(E_{\pi})}{\sqrt{E_{\pi}^{2}- m_{\pi}^{2} c^4}}
\;dE_{\pi}, \label{qgama}
\end{equation}  
where $E_{\pi}^{\rm min}(E_{\gamma})=E_{\gamma}+{m_{\pi}^{2}
c^4}/{4E_{\gamma}}$.
Applying the $\delta$-functional approximation for
the differential cross section\footnote{This approximation considers
only the most energetic neutral pion that is produced in the $pp$
reaction aside of a {\em fireball} composed by a certain number
of less energetic $\pi$-mesons of each flavor. See
a discussion in \citet{Pfrommer04}.}
\citep{Aharonian00}, the pion emissivity becomes
\begin{eqnarray}\label{q_pi}
q_{\pi^0}(E_{\pi}) &=\frac{4\pi}{\kappa} \, n_{\rm H}\,J_p\left( m_p
c^2+\frac{E_{\pi}}{\kappa}\right)\sigma_{pp}\left( m_p
c^2+\frac{E_{\pi}}{\kappa}\right)
\end{eqnarray}
for proton energies greater than the energy threshold $E_{\rm
th}=1.22$ GeV and lower than $100$ GeV. 
Here, $\kappa$ is the mean fraction of the kinetic
energy $E_{\rm kin}=E_p-m_p c^2$ of the proton transferred to a
secondary meson per collision. For a broad energy region (GeV to TeV)
we have that $\kappa\sim 0.17$. 
The proton flux is given by $J_p(E_p) = (4\pi/c)\,n(E_p)$. 
The total cross section of the inelastic $pp$
collisions is  given by Eq.~(\ref{cross-section}). 
For $E_p > 100$ GeV, the equations given by \citet{kelner06} 
are used.
The specific luminosity is estimated in the same way as for electrons
(Eq.~\ref{luminosity}).

\begin{figure}
\includegraphics[width=0.9\hsize]{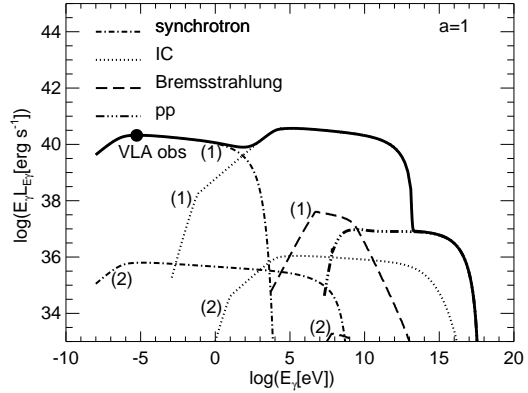} %
\caption{SED for a mixed primary population of relativistic electrons and 
protons characterized by $a=1$, represented by a solid line.
The contribution of different radiation processes are identified
as in Fig.~\ref{SED_0}. 
The contribution from secondary pairs is also included.}
\label{SED_1}
\end{figure}

The results of our calculations for the case $a=1$ are shown in 
Fig.~\ref{SED_1}. 
Just as the case charaterized by $a=0$, the SED is dominated
by the IC interactions, with a luminosity 
$L_{\rm IC} \sim 7.4\times 10^{41}$ erg s$^{-1}$. The emission
at  energies higher than $\sim 1$~GeV is produced by neutral pion decay, 
reaching a luminosity
$L_{pp} \sim 1.6\times 10^{38}$ erg s$^{-1}$ with a cut-off  at 
$E_{\gamma} \sim 10^{17}$ eV. However, its contribution to the SED
becomes evident at energies $\gtrsim 10$~TeV. 

\begin{figure}
\includegraphics[width=0.9\hsize]{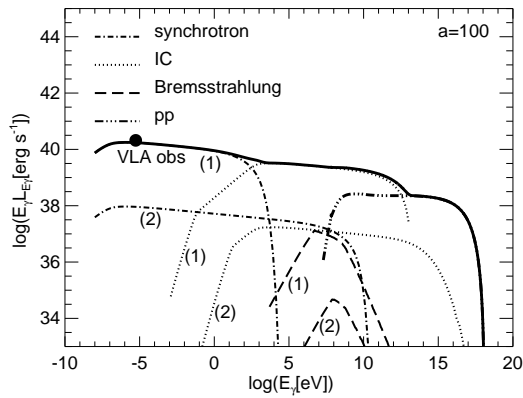} %
\caption{Idem as Fig~\ref{SED_1}, but for a case dominated by relativistic 
protons ($a=100$).}
\label{SED_100}
\end{figure}

In the case corresponding to $a = 100$, the
relativistic energy density of protons is higher than that of the  
primary electrons.
However, the proton dominance 
does not imply necessarily a significant increase of photon production 
from $pp$-interactions because of the low density of thermal protons
at the location of radio relics.
Hence,  
the power produced by $\pi^0$-decay does not
dominate the SEDs, 
as can be seen in Fig.~\ref{SED_100}. The luminosity $L_{pp}$
is of the order of $\sim 4.2\times10^{39} \, {\rm erg}\,{\rm s}^{-1}$, being 
slightly larger than the one corresponding to the case with $a=1$. 
In contrast to the cases with $a=0$ and $a=1$, 
the IC emission is lower than the synchrotron one, with luminosities 
$L_{\rm IC} \sim 7.1\times10^{40}\,{\rm erg}\, {\rm s}^{-1}$ and 
$L_{\rm synch} \sim 3.8\times10^{41}\,{\rm erg}\, {\rm s}^{-1}$, respectively.
The fact that radiation produced by IC scattering is reduced
for the case $a=100$ might be explained as a consequence of
equipartition  between the magnetic field and the particle energy density.
As can be seen in Table~\ref{table1}, the magnetic field becomes larger as
the value of $a$ increases.  
This will reduce the amount of electron energy released via IC scattering
with respect to that released via synchrotron. In addition, the available
electron energy is also reduced to explain the radio fluxes due to the
larger magnetic field. All this makes the IC luminosity smaller.
Concerning the emission produced by secondary pairs, we find
that their contributions to the SEDs by different radiation processes 
are larger than that for the
case $a=1$, but are still irrelevant. This is consistent with 
the low efficiency of $pp$-interactions taking place
in the outskirts of the cluster.

\subsection{Gamma-ray emission}

The SED shown in Figs.~\ref{SED_0} to \ref{SED_100} 
are built considering the  contribution of different non-thermal radiative
processes in a wide range of energies, from radio to gamma-rays.
In the present work, we are particularly interested in the
detectability of the cluster Abell 3376 at gamma-ray energies.
We find that the main processes contributing to the emission
at these high energies are IC scattering and $\pi^0$-decay,
being the luminosity of the latter considerably smaller
even for the case where a large energy density of protons
is considered. 
The HESS array could detect the estimated emission from  A3376 in 
exposure times between one day to one month, for different values of $a$. 
On the other hand, more than one year is necessary to detect with GLAST 
the non-thermal gamma-ray emission
estimated for the cluster studied in this work. However, small changes 
in the assumptions of our model can yield higher fluxes that would result 
in a considerably reduction of the exposure time
for detection. For instance, if subpartition between magnetic and relativistic
particle energy densities is considered instead of equipartition, there 
would be an increase of the IC emission.
Thus, if the IC luminosity of Abell 3376 is 
$L_{\rm IC} \sim 10^{42}$~erg~s$^{-1}$ at energies above 100~MeV it would 
be detected by GLAST in less than one year.  
(i.e. within the initial all-sky survey). Finally,
photons produced by $pp$-interactions could be might by the
future Cherenkov telescope HESS II on short timescales.

\section{Summary and prospects}
\label{s_prospects}

We have presented a model for the emission produced by the nearby cluster
Abell 3376 
in a wide range of energies, from radio to gamma-rays,
considering the contribution of different non-thermal radiative
processes.
This cluster presents strong evidence of merger activity, 
characterized by large-scale ring-shaped synchrotron
radio structures, that are identified with radio relics \citep{Bagchi06}.
These kind of radio structures detected in several clusters of galaxies 
suggest 
a rich content of accelerated particles in the ICM, with
energies as high as  $\sim 10^{14}$ and $\sim 10^{18}$ eV for electrons and 
protons, respectively. 
Taking into account that radio relics are tracers of 
merger shocks, we assume that a
diffusive shock acceleration mechanism acting at the 
wave shocks gives place to such a population of relativistic particles.
These particles are subsequently cooled by different
leptonic and hadronic radiative processes. 

We have considered syncrotron radiation, IC scattering and relativistic 
Bremsstrahlung that affect relativistic primary electrons.
Relativistic protons are involved in $pp$-interactions with the thermal
protons of the ICM. This process give raise to a population of secondary
pairs, $e^{\pm}$, that are cooled by the same processes that 
affect the population of primary electrons. 
The parameters involved in these
non-thermal processes, such as the magnetic field and normalization
of the energy distribution of accelerated particles,
have been estimated from the constraints established
by the observed radio power in the relics and the assumption
of equipartition.
X-ray observations allows us to extract information about the
volume of these acceleration sources. 
On the other hand, results from a simulated cluster have provided
the typical gas density of the relics and the velocity of ths shock.

We find that particle energy distributions reach a stationary regime
by the lifetime of the relics ($\sim 1$~Gyr).
The steady energy distribution of particles are computed 
and the SEDs of the radiation produced at the relic position derived.
In our calculations, we have adopted different ratios ($a = 0, 1, 100$)
for the relativistic proton/electron energy densities.
At larger energies, gamma-ray emission is mainly produced by
$\pi^0$-decay but with a considerably smaller luminosity, 
even for the case where a large energy density of protons
is considered ($L_{pp} \sim 4\times 10^{39}$ erg s$^{-1}$).
This is consistent with the fact that 
no cluster has been yet observed at these high energies.

The proximity of the cluster studied in this paper and its high
content of relativistic particles in the radio-relics 
make Abell 3376 an interesting potential target 
for the investigation of gamma-ray emission in this type of objects. 
This source might be detected at
$\gamma$-rays by GLAST satellite and by HESS array 
with a reasonable exposure time. In addition, future
Cherenkov telescopes in the southern hemisphere, like HESS II, 
could easily detect the high-energy emission from Abell 3376.

\section*{Acknowledgments}

We are grateful to an anonymous referee for insightful remarks. We thank 
Valent{\'\i} Bosch-Ramon for helful discussion and 
comments about this work. Marcus Br\"ueggen is acknowledged for
useful remarks.  
A.T.A. and G.E.R. are supported by CONICET (PIP 5375) and the
Argentine agency ANPCyT through Grant PICT 03-13291 BID 1728/OC-AC.
S.A.C is supported by CONICET (PIP 5000/2005)
and the Argentine agency ANPCyT through Grant 
PICT 26049 BID 1728/OC-AC.


%


\begin{thebibliography}{}

%
\bibitem[\protect\citeauthoryear{Aharonian \& Atoyan}{2000}]{Aharonian00}
Aharonian, F.A., Atoyan A.M., 2000, A\&A, 362, 937
\bibitem[\protect\citeauthoryear{Atoyan \& V\"olk}{2000}]{AtoyanVolk00}
Atoyan A.M., V\"olk H.J., 2000, ApJ, 535, 45
\bibitem[\protect\citeauthoryear{Bagchi et al.}{2006}]{Bagchi06}
Bagchi J., Durret F., Neto G.B.L., Paul S., 2006, Sci, 314, 791
\bibitem[\protect\citeauthoryear{Balestra et al.}{2007}]{balestra07}
Balestra I., Tozzi P., Ettori S., Rosati P., Borgani S., Mainieri V., 
Norman C., Viola M., 2007, A\&A, 462, 429 
\bibitem[\protect\citeauthoryear{Berezinsky \& Grigoreva}{1988}]{Berezinsky88}
Berezinsky V.S. \& Grigorieva S.I., 1988, A\&A, 199, 1
\bibitem[\protect\citeauthoryear{Berrington \& Dermer}{2003}]
{BerringtonDermer03}
Berrington, R.C. \& Dermer C.D., 2003, ApJ, 594, 709
\bibitem[\protect\citeauthoryear{Blasi}{2003}]{Blasi}
Blasi P., 2003, ASP Conference Proceedings, 301, 203 
\bibitem[\protect\citeauthoryear{Blumenthal \& Gould}{1970}]{Blumenthal70}
Blumenthal G.R., Gould R.J., 1970, Rev. Mod. Phys., 42, 237
\bibitem[\protect\citeauthoryear{Bowyer et al.}{2004}]{Bowyer04}
Bowyer S., Korpela E.J., Lampton M., Jones T.W., 2004, ApJ, 605, 168
\bibitem[\protect\citeauthoryear{Hoeft et al.}{2004}]{Hoeft04}
Hoeft M., Br\"ueggen M., Yepes G. 2004, MNRAS, 347, 389
\bibitem[\protect\citeauthoryear{Churazov et al.}{2000}]{Churazov00}
Churazov E., Forman W., Jones C., B\"ohringer H., 2000, A\&A, 356, 788
\bibitem[\protect\citeauthoryear{Dolag et al.}{2005}]{Dolag05}
Dolag K., Vazza F., Brunetti G., Tormen G., 2005, MNRAS, 354, 753
\bibitem[\protect\citeauthoryear{Domainko et al.}{2007}]{Domainko07}
Domainko W., Benbow W., Hinton J.A., Martineau-Huynh O., de Naurois M., 
Nedbal D., Pedaletti G., Rowell G., for the H. E. S. S. Collaboration
2007, 30th International
Cosmic Ray Conference, Merida, Mexico, astro-ph/0708.1384v1
\bibitem[\protect\citeauthoryear{Drury}{1983}]{Drury83}
Drury L.O., 1983, Rep. Prog. Phys., 46, 973
\bibitem[\protect\citeauthoryear{En$\beta$lin \& Biermann}{1998}]{EnsslinBiermann98}
En$\beta$lin T. A., Biermann P. L., 1998, A\&A, 330, 90
\bibitem[\protect\citeauthoryear{En$\beta$lin et al.}{1998}]{Ensslin98}
En$\rm{\beta}$lin T. A., Biermann P. L., Klein U., Kohle S. 1998,
A\&A, 332, 395 
\bibitem[\protect\citeauthoryear{En$\beta$lin \& Gopal-Krishna}{2001}]{EnsslinGopal01}
En$\rm \beta$lin T. A., Gopal-Krishna, 2001, A\&A, 366, 26
\bibitem[\protect\citeauthoryear{Fegan et al.}{2005}]{Fegan05}
Fegan S.J., Badran H.M., Bond I.H., Boyle P.J., Bradbury S.M., Buckley J.H., Carter-Lewis D.A., Catanese M., et al., 2005, ApJ, 624, 638
\bibitem[\protect\citeauthoryear{Feretti \& Giovannini}{1996}]{Feretti96}
Feretti L., Givannini G., 1996. In R. Ekers, C. Fanti \^ L. Padrielli (eds.)
 IAU Symp. 175, Extragalactic Radio Sources. Kluwer Academic Publisher, p. 333 
\bibitem[\protect\citeauthoryear{Feretti et al.}{1997}]{Feretti97}
Feretti L., B\"ohringer H., Giovannini G., Neumann D., 1997, A\&A, 317, 432
\bibitem[\protect\citeauthoryear{Feretti et al.}{2001}]{Feretti01}
Feretti L., Fusco-Femiano R., Giovannini G., Govoni F., 2001, A\&A, 373, 106
\bibitem[\protect\citeauthoryear{Feretti, Burigana \& En$\rm \beta$lin}{2004}]{Feretti04}
Feretti L., Burigana C., En$\beta$lin T.A., 2004, New Astron. Rev., 48, 1137
\bibitem[\protect\citeauthoryear{Feretti \& Giovannini}{2008}]{FerettiGiovannini08}
Feretti L., Giovannini G., 2008, in  Plionis M., Lopez-Cruz O., Hughes D. (eds.)
 Panchromatic View of Clusters of Galaxies and the Large-Scale Structure, 
Lecture Notes Physics 740, Springer, Dordrecht, p. 143
\bibitem[\protect\citeauthoryear{Fusco-Femiano et al.}{1999}]{FuscoFemiano99}
Fusco-Femiano R., dal Fiume D., Feretti L., Giovannini G., Grandi P., Matt G.,
Molendi S., Santangelo A., 1999, ApJ, 513, L21
\bibitem[\protect\citeauthoryear{Fusco-Femiano et al.}{2004}]{fuscofemiano04}
Fusco-Femiano R., Orlandini M., Brunetti G., Feretti L., Giovannini G., Grandi P., Setti G., 2004, ApJ, 602, L73
\bibitem[\protect\citeauthoryear{Gabici \& Blasi}{2003}]{Gabici}
Gabici S., Blasi P., 2003, ApJ, 583, 695
\bibitem[\protect\citeauthoryear{Ginzburg \& Syrovatskii}{1964}]{GZ}
Ginzburg, V.L. \& Syrovatskii, S.I., 1964, The Origin of Cosmic Rays, Pergamon Press, New York
\bibitem[\protect\citeauthoryear{Giovannini et al.}{1991}]{Giovannini91}
Giovannini G. Feretti L., Stanghellini C., 1991, A\&A, 252, 528
\bibitem[\protect\citeauthoryear{Giovannini et al.}{1999}]{Giovannini99}
Giovannini G., Tordi M., Feretti L., 1999, New Astron., 4, 141
\bibitem[\protect\citeauthoryear{Giovannini \& Ferreti}{2004}]{GiovanniniFerreti04}
Giovanninii G., Ferreti L., 2004, JKAS, 37, 323
\bibitem[\protect\citeauthoryear{Girardi et. al}{1998}]{Girardi98}
Girardi M., Giuricin G., Mardirossian F., Mezzetti M., Boschin W. 1998, ApJ, 505, 74
\bibitem[\protect\citeauthoryear{Govoni et al.}{2001}]{Govoni01}
Govoni F, Feretti L., Giovannini G., B\"ohringer H., Reiprich T.H., Murgia M. 2001, A\&A, 376, 803 
\bibitem[\protect\citeauthoryear{Govoni \& Feretti}{2004}]{GovoniFeretti04}
Govoni F, Feretti L., 2004, Int. J. Mod. Phys. D, 13, 1549
\bibitem[\protect\citeauthoryear{Hoeft et al.}{2004}]{Hoeft04}
Hoeft M., Br\"ueggen M., Yepes G. 2004, MNRAS, 347, 389
\bibitem[\protect\citeauthoryear{Jaffe}{1977}]{Jaffe77}
Jaffe W.J., 1977, ApJ, 212, 1
\bibitem[\protect\citeauthoryear{Khangulyan et al.}{2007}]{Khangulyan07}
Khangulyan D., Hnatic S., Aharonian F., Bogovalov S., S2007, MNRAS,380, 312 
\bibitem[\protect\citeauthoryear{Kelner et al.}{2006}]{kelner06}
Kelner, S.R., Aharonian, F.A., \& Vugayov, V.V., 2006, Phys. Rev. D, 74, 034018
\bibitem[\protect\citeauthoryear{Kelner \& Aharonian}{2008}]{kelner08}
Kelner, S.R., \& Aharonian, F.A., 2008 (submitted) [arXiv:0803.0688v1]
\bibitem[\protect\citeauthoryear{Keshet}{2003}]{Keshet03}
Keshet U., Waxman E., Loeb A., Springel V., Hernquist L., 2003, ApJ, 585, 128
\bibitem[\protect\citeauthoryear{Lieu et al.}{1996}]{Lieu96}
Lieu R., Mittaz J.P.D., Bowyer S., Lockman F., Hwang C.-Y., Schmitt J.H.M.M.,
1996, ApJ, 458, L5
\bibitem[\protect\citeauthoryear{Lieu et al.}{1999}]{Lieu99}
Lieu R., Axford W. I., Bonamente M., 1999, ApJ, 510, L25
\bibitem[\protect\citeauthoryear{Mannheim \& Schlickeiser}{1994}]{Mannheim94}
Mannheim K. \& Schlickeiser R., 1994, A\&A, 286, 983
\bibitem[\protect\citeauthoryear{Mittaz}{1998}]{Mittaz98}
Mittaz J.P.D., Lieu R., Lockman F.J., 1998, ApJ, 498, L17
\bibitem[\protect\citeauthoryear{Pacholczyk}{1970}]{Pacholczyk70}
Pacholczyk, A.G., 1970, Radio Astrophysics, Freeman, San Francisco
\bibitem[\protect\citeauthoryear{Perkins}{2006}]{Perkins06}
Perkins J.S., Badran H.M., Blaylock G., Bradbury S.M., Cogan P., Chow Y.C.K., 
Cui W., Daniel M.K., et al., 2006, ApJ, 644, 148 
\bibitem[\protect\citeauthoryear{Petrosian}{2001}]{Petrosian01}
Petrosian V., 2001, ApJ, 557, 560
\bibitem[\protect\citeauthoryear{Pfrommer \& En$\beta$lin}{2004}]{Pfrommer04}
Pfrommer C., En$\beta$lin T.A., 2004, A\&A, 413, 17
\bibitem[\protect\citeauthoryear{Pfrommer et al.}{2006}]{Pfrommer06}
Pfrommer C., Springel V., En$\beta$lin T. A., Jubelgas M., 2006, MNRAS, 367, 113
\bibitem[\protect\citeauthoryear{Pfrommer et al.}{2007a}]{Pfrommer07}
Pfrommer C., En$\beta$lin T. A., Springel V., Jubelgas M., Dolag K., 2007, MNRAS, 378, 385
\bibitem[\protect\citeauthoryear{Pfrommer et al.}{2008}]{Pfrommer08}
Pfrommer C., En$\beta$lin T. A., Springel V., 2008, MNRAS, 385, 1211
\bibitem[\protect\citeauthoryear{Reimer et al.}{2003}]{Reimer03}
Reimer O., Pohl M., Sreekumar P., Mattox J. R., 2003, ApJ, 588, 155
\bibitem[\protect\citeauthoryear{R\"ottgering et al.}{1994}]{Rottgering94}
R\"ottgering H.J.A., Snellen I., Miley G., de Jong J.P., Hanisch R.J., Perley R., 1994, ApJ, 436, 654 
\bibitem[\protect\citeauthoryear{R\"ottgering et al.}{1997}]{Rottgering97}
R\"ottgering H.J.A., Wieringa M.H., Hunstead R.W., Ekers R.D., 1997, MNRAS, 
290, 577
\bibitem[\protect\citeauthoryear{V\"olk, Aharonian \& Breitschwerdt}{1996}]{VolkAha96}
V\"olk H.J., Aharonian F.A., Breitschwerdt D., 1996, SSRv, 75, 279

\end{thebibliography}
\end{document}